\documentclass[prd,superscriptaddress,floatfix,amsfonts,amssymb,amsmath,showpacs,twocolumn]{revtex4-2}
\usepackage{bm}
\usepackage{amsfonts}
\usepackage{latexsym}
\usepackage[latin1]{inputenc}
\usepackage{graphicx}
\usepackage{amsmath}
\usepackage{palatino}
\usepackage{mathpazo}
\usepackage{textcomp}
\usepackage{booktabs}
\usepackage{dcolumn}
\usepackage{hyperref}
\hypersetup{colorlinks=true,citecolor=blue,urlcolor=blue,anchorcolor = blue}
\usepackage{amsmath}
\usepackage{xcolor}
\usepackage{mathtools}
\usepackage{multirow}
\usepackage{float}
\usepackage{orcidlink}

\allowdisplaybreaks[1]

\begin{document}
\color{red}

\title{Bulk viscous matter and the cosmic acceleration of the universe in $f(Q,T)$ gravity}

\author{Simran Arora\orcidlink{0000-0003-0326-8945}}
 \email{dawrasimran27@gmail.com}
\affiliation{ Department of Mathematics, Birla Institute of Technology and Science-Pilani,\\ Hyderabad Campus, Hyderabad-500078, India}
\author{S. K. J. Pacif\orcidlink{0000-0003-0951-414X}}
\email{shibesh.math@gmail.com}
\affiliation{Centre for Cosmology and Science Popularization (CCSP) SGT University, Gurugram, Delhi-NCR, Haryana-122505, India}
\author{Abhishek Parida}
 \email{abhishekparida22@gmail.com}
\affiliation{International College of Liberal Arts, Yamanashi Gakuin University, Yamanashi 400-0805, Japan}
\author{P.K. Sahoo\orcidlink{0000-0003-2130-8832}}
 \email{pksahoo@hyderabad.bits-pilani.ac.in}
\affiliation{ Department of Mathematics, Birla Institute of Technology and Science-Pilani,\\ Hyderabad Campus, Hyderabad-500078, India}
\date{\today}

\begin{abstract}
We have studied bulk viscosity in the modified $f(Q, T)$ gravity theory formalism, where $Q$ represents the non-metricity and $T$ denotes the trace of energy-momentum tensor within a flat Friedmann-Lema\^{i}tre-Robertson-Walker metric (FLRW). Here, we have explicitly considered the effective equation of state, which includes a bulk viscosity term, and obtained the exact solutions by assuming a specific form of $f(Q, T)=\alpha Q+\beta T$, where $\alpha$ and $\beta$ are constants. Furthermore, we have found constraints on the model parameters with some external datasets, such as the revised Hubble datasets consisting of 57 data points, Baryon acoustic oscillations (BAO) datasets, and the newly published Pantheon samples with 1048 points to obtain the best fitting values of the model parameters. The obtained model is found to be in good agreement with observations. In addition, we have analyzed the cosmological behavior of the density parameter, the equation of state (EoS) parameter ($\omega$), and the deceleration parameter ($q$).  The results are satisfying to the standard scenario of recent findings of cosmology. The universe appears to be evolving from a decelerated to an accelerated phase. The EoS parameter is further in the quintessence phase, indicating that the universe is accelerating. Finally, we can deduce that the accumulation of bulk viscosity as effective dark energy supports the current accelerated expansion of the universe.
\end{abstract}

\keywords{$f(Q,T)$ gravity; Bulk viscosity, Dark energy, Observational constraints}
\pacs{04.50.Kd}
\maketitle

\section{Introduction}\label{sec1} 
Observations of type Ia supernovae in recent years have revealed that the universe is expanding at a higher pace than usual \cite{Riess/1998,Perlmutter/1999}. To understand the acceleration, physicists proposed a new fluid termed dark energy, which has enough negative pressure. It appears to be seventy percent of the total content of the energy and matter in the universe. The straightforward way to explain dark energy is to introduce a cosmological constant (CC), which leads to the alleged accelerated expansion, and establishes the $\Lambda$CDM model, which has proven to be accurate \cite{Zlatev/1999}. Even though observations favor the cosmological constant model, it has several serious flaws. The significant difference between its expected and observed value of the cosmological constant and the cosmic coincidence is distinguished by the fact that we live specifically in the universe whose matter density and dark energy density are of the same order. This encourages researchers to investigate alternative models for accelerated expansion \cite{Weinberg/1998, Pad/2003,Steinhardt/1999}. 

Modified gravity has recently emerged as an influential branch of modern cosmology, seeking to provide a coherent explanation of the early epoch of the universe while also accounting for the accelerated expansion at later stages. Modified gravity theories are geometrical extensions of Einstein General Relativity (GR), in which Einstein-Hilbert action is altered to accomplish cosmic acceleration. Many modified gravity theories have been proposed to describe the late and early acceleration of the universe. 
The $f(R)$ gravity proposed in \cite{Buchdahl/1970}, is the most fundamental and widely used modification to GR. Several authors have investigated several aspects of $f(R)$ gravity and how it can cause cosmic inflation and acceleration \cite{Dunsby/2010,Carroll/2004}. The existence of the non-minimal coupling between matter and geometry is another extension of Einstein-Hilbert's action. As a result, this leads to the so-called $f(R,T)$ modified theory of gravity. Harko et al. \cite{Harko/2011} proposed $f(R,T)$ gravity, in which the gravitational Lagrangian is described by an arbitrary function of Ricci scalar $R$ and the trace of energy-momentum tensor $T$. There are some astrophysical and cosmological implications investigated in $f(R,T)$ gravity \cite{Yousaf/2016}. Harko \cite{THarko/2014} presented the thermodynamic interpretation of generalized gravity models with geometry-matter coupling. Jamil et al. \cite{Jamil/2012} also reconstructed several cosmological models in $f(R,T)$ theory. Moraes and Sahoo \cite{Moraes/2017} proposed the modeling of static wormholes within the $f(R,T)$ extended theory of gravity. Similarly, there exists several other modified theories with different cosmological implications such as $f(G)$ theory \cite{Felice/2009,Bamba/2017,Goheer/2009}, $f(R, G)$ theory \cite{Elizalde/2010,Bamba/2010}, $f(T,B)$ theory \cite{Bahamonde/2018}, etc.

The GR theory is known to be expressed in Riemannian geometry. Hence, investigating more general geometric structures that could characterize the gravitational field at the solar system level is another intriguing method for developing extended theories of gravity. As a result, a more unified theory equivalent to GR known as the teleparallel equivalent to GR or $f(T)$ theory \cite{Ferraro/2007,Myrzakulov/2011,Capozziello/2011} has been developed, where $T$ is the torsion describing the gravitational effects. Further, Weyl proposed an extension of Riemannian geometry, in which he established the first unified theory of gravity and electromagnetism, where the non-metricity of spacetime generated the electromagnetic field. As a result, the symmetric teleparallel representation is the third generalization of GR.  This generalization is extended to the non-metric gravity known as the $f(Q)$ gravity \cite{Jimenez/2018,Jimenez/2020}, where $Q$ is the non-metricity representing the geometric variable defining the characteristics of the gravitational interaction. Studies on the cosmology of the $f(Q)$ theory with observational constraints demonstrate the accelerated expansion of the universe without the need for unusual dark energy or other fields \cite{Lazkoz/2019,Lu/2019}. Mandal et al. \cite{Mandal/2020} investigated energy conditions in $f(Q)$ theory and compared the study to $\Lambda$CDM. Khyllep et al. \cite{Khyllep/2021} examined cosmological solutions and the growth index of matter perturbations in $f(Q)$ gravity.  
Recently, Yixin et al. \cite{Yixin/2019,Yixin/2020} have proposed a new extension of $f(Q)$ known as the $f(Q, T)$ theory, where the non-metricity $Q$ is non-minimally coupled to the trace $T$ of the energy-momentum tensor. The fundamental quantity in the $f(Q, T)$ gravity is still a metric that explains the fundamental aspects of gravitational interaction and geometric description. It is seen that for $T=0$ ( the case of vacuum), the theory reduces to the $f(Q)$ gravity, which is equivalent to GR and passes all solar system tests. The $f(Q, T)$ theory coupling, like the standard curvature trace of the energy-momentum tensor couplings, leads to non-conservation of the energy-momentum tensor. This violation of conservation has substantial physical indications predicting important changes in the thermodynamics of the universe comparable to those anticipated by the modified $f(R, T)$ gravity. Yixin et al. \cite{Yixin/2019,Yixin/2020} focused on the three types of basic models and came up with solutions that describe both the accelerated and decelerated evolutionary phases of the universe. Arora et al. \cite{Arora/2020} also tested $f(Q,T)$ gravity models with observational constraints to address the present cosmic acceleration. Yang et al. \cite{Yang/2021} developed the geodesic deviation and Raychaudhuri equations in the Weyl-type $f(Q, T)$ gravity based on the observation that the curvature-matter coupling considerably modifies the nature of tidal forces and the equation of motion in the Newtonian limit. Therefore, one can study the viability of this newly proposed $f(Q, T)$ gravity under different cosmological implications. \\
Since the introduction of relativistic thermodynamics, viscous cosmological scenarios have been extensively investigated. In 1940, Eckart obtained the standard expression for relativistic viscosity \cite{Eckart/1940}. Moreover, Treciokas and Ellis \cite{Treciokas/1971}, and Weinberg \cite{Weinberg/1971} examined the cosmological implications of the Eckart viscosity. The concept of bulk viscosity was later explored in the form of inflation \cite{Pady/1987, Barrow/1987} and as a source for the rapid expansion of the universe \cite{Velten/2013}. The bulk viscosity in a cosmic fluid can occur when it expands faster than the system has time to recover its local thermodynamic equilibrium. An effective pressure emerges, restoring the system to its thermal stability \cite{Ilg/1999}. This effective pressure can be regarded as an indicator of bulk viscosity. Perhaps, due to the presence of bulk viscosity, it is reasonable to assume that the expansion phase is simply a set of states out of thermal equilibrium in a limited fraction of time in an accelerated expanding universe \cite{Wilson/2007}. There have been theories about viscous fluids playing a part in dark matter \cite{Velte/2012} and dark energy in the literature \cite{Gagnon/2011,Cataldo/2005}. The influence of bulk viscosity has been examined in the context of the late acceleration of the universe in literature \cite{Y/2012,Feng/2009,SCapo/2006,Gagnon/2011,Mohan/2017}. In Ref. \cite{Mak/1998}, the authors established exact solutions that correspond to the early inflationary period of the universe with a bulk viscous coefficient proportionate to the Hubble parameter, which is an exciting study of the cosmology of flat FLRW viscous universe. Concern over the bulk viscosity is understandable and practical at late times as we do not know the nature of the contents of the universe i.e, dark energy and dark matter. Such a notion has only been considered in the context of non-singular model searches and the primordial universe. So, this study aims to use the bulk viscous within the cosmic fluid instead of any dark energy component in modified gravity to drive the current acceleration.\\
The CC is the simplest model to study dark energy, whereas issues such as appropriate regulation and the challenge of cosmic adaptability urge us to seek another explanation for the expansion of the universe. The modification of the geometric section of the general relativity field equations is the other option, which has lately been examined. However, several studies suggest that viscous pressure may be a driving force behind the current acceleration of the universe \cite{Avelino/2010,Ren/2006,Disconzi/2015}. Srivastava and Singh \cite{Srivastava/2018} studied new holographic dark energy (HDE) model in modified $f(R, T)$ gravity theory within the framework of a FLRW model with bulk viscous matter content (considering $p_{eff}=p-3\zeta H$, $\zeta$ as constant). Brevik et al. \cite{Brevik/2005,Brevik/2012} also discussed viscous FRW cosmology in modified gravity. Singh and Kumar \cite{Singh/2014} introduced bulk viscosity in modified $f(R, T)$ gravity with viscous term as $\zeta = \zeta_{0} + \zeta_{1}H$, where  $\zeta_{0}$,  $\zeta_{1}$ are constants and $H$ is the Hubble parameter. Davood \cite{Davood/2019} studied the effect of bulk viscosity matter in $f(T)$ gravity.  Arora et al. \cite{Simran/2020} used an effective equation of state to investigate cosmological evolution with bulk viscosity in $f(R, T)$ theory. \\
This work aims to use the bulk viscous pressure inside the cosmic fluid without incorporating any dark energy element in the $f(Q, T)$ modified theory of gravity to drive the current acceleration. We consider the bulk viscosity of the form $\zeta= \zeta_{0} + \zeta_{1} H$, where $H$ is the Hubble parameter, $\zeta_{0}$, and $\zeta_{1}$ are constants. The term $\zeta_{0}$ refers to the most basic parametrization of bulk viscosity, which is a constant and $\zeta_{1}$ refers to the possibility of a bulk viscosity proportional to the expansion rate of the universe. Also, the functional form of $f(Q, T)= \alpha Q+ \beta T $ is assumed to get the exact solutions of field equations. We try to constrain the model parameters using the released 57 Hubble data points and 1048 Pantheon samples. To find the optimal values for the model parameters, we use MCMC techniques.\\
The following article is organized into sections. The field equation formalism in $f(Q, T)$ gravity is presented in section \ref{sec2}. In section \ref{sec3}, we presented the FLRW universe dominated by bulk viscous matter and obtained the Hubble parameter expression. We used the revised 57 Hubble datasets and Pantheon samples to constrain the model parameters in section \ref{sec4}. We observed the behavior of cosmological parameters such as the density parameter, the EoS parameter, and the deceleration parameter in section \ref{sec5}. Finally, in section \ref{sec6}, we discussed our conclusions.

\section{Basics of $f(Q,T)$ Gravity}\label{sec2}

The action used to define $f(Q,T)$ gravity read as \cite{Yixin/2019,
Yixin/2020},

\begin{equation}  \label{1}
S=\int \left( \frac{1}{16\pi} f(Q,T) + L_{m}\right )\sqrt{-g} d^4x,
\end{equation}
where $f(Q,T)$ is an arbitrary function that couples the non-metricity $Q$
and $T$, the trace of the energy-momentum tensor. Further, $L_{m}$ serve as
the matter Lagrangian and $g = det(g_{\mu \nu})$. The non-metricity $Q$ is defined as \cite{Jimenez/2018}

\begin{equation}  \label{2}
Q\equiv -g^{\mu \nu}(L^{ \beta}_{\,\,\, \alpha \mu}L^{\alpha}_{\,\,\, \nu
\beta}-L^{\beta}_{\,\,\, \alpha \beta}L^{\alpha}_{\,\,\, \mu \nu}),
\end{equation}
where the disformation tensor $L^{\beta}_{\,\,\, \alpha\gamma}$ is written
as, 
\begin{equation}  \label{3}
L^{\beta}_{\alpha\gamma}=-\frac{1}{2}g^{\beta\eta}(\nabla_{\gamma}g_{\alpha
\eta}+\nabla_{\alpha}g_{\eta\gamma}-\nabla_{\eta}g_{\alpha \gamma}).
\end{equation}

The non-metricity tensor is defined by 
\begin{equation}  \label{4}
Q_{\gamma\mu\nu}=\nabla_{\gamma}g_{\mu\nu},
\end{equation}
and trace of the non-metricity tensor is obtained as follows. 
\begin{equation}  \label{5}
Q_{\beta}= g^{\mu \nu}Q_{\beta \mu \nu} \qquad \widetilde{Q}_{\beta}= g^{\mu
\nu}Q_{\mu \beta \nu}.
\end{equation}
We can also define a superpotential or the non-metricity conjugate as 
\begin{equation}  \label{6}
P^{\beta}_{\,\,\, \mu \nu}= -\frac{1}{2} L^{\beta}_{\,\,\, \mu \nu}+ \frac{1%
}{4} (Q^{\beta}- \widetilde{Q}^{\beta})g_{\mu \nu} - \frac{1}{4}
\delta^{\beta}_{(\mu}Q_{\nu)}.
\end{equation}
giving the non-metricity scalar as \cite{Jimenez/2018} 
\begin{equation}  \label{7}
Q=-Q_{\beta\mu\nu}P^{\beta\mu\nu}\,.
\end{equation}

Besides, the energy-momentum tensor is defined as 
\begin{equation}  \label{8}
T_{\mu \nu}= -\frac{2}{\sqrt{-g}} \dfrac{\delta(\sqrt{-g}L_{m})}{\delta
g^{\mu \nu}},
\end{equation}
and 
\begin{equation}  \label{9}
\Theta_{\mu \nu}= g^{\alpha \beta} \frac{\delta T_{\alpha \beta}}{\delta
g^{\mu \nu}}.
\end{equation}

Further, the variation of energy-momentum tensor with respect to the metric
tensor read as 
\begin{equation}  \label{10}
\frac{\delta\,g^{\,\mu\nu}\,T_{\,\mu\nu}}{\delta\,g^{\,\alpha\,\beta}}=
T_{\,\mu \nu}+\Theta_{\,\mu \nu}\,.
\end{equation}

Hence, the following field equations is obtained after varying the action \eqref{1} with respect to the metric and equating it to zero. 
\begin{multline}  \label{11}
-\frac{2}{\sqrt{-g}}\nabla_{\beta}(f_{Q}\sqrt{-g} P^{\beta}_{\,\,\,\, \mu
\nu}-\frac{1}{2}f g_{\mu \nu}+ f_{T}(T_{\mu \nu}+\Theta_{\mu \nu}) \\
-f_{Q}(P_{\mu \beta \alpha}Q_{\nu}^{\,\,\, \beta \alpha}-2Q^{\beta
\alpha}_{\, \, \, \mu}P_{\beta \alpha\nu})= 8\pi T_{\mu \nu},
\end{multline}
where $f_{Q}= \dfrac{df}{dQ}$ and $f_{T}= \dfrac{df}{dT}$.

\section{Friedmann cosmology with bulk viscosity}\label{sec3}

We assume the bulk viscosity coefficient as \cite{Meng/2009} 
\begin{equation}  \label{12}
\zeta= \zeta_{0}+\zeta_{1} H,
\end{equation}

where $\zeta_{0}$ and $\zeta_{1}$ are two constants, regarded as positive
and $H= \frac{\dot{a}}{a}$. Here, dot($\cdot$) represents the derivative
with respect to time. The reason for addressing this bulk viscosity is since
we know the transport viscosity phenomenon are related to the velocity $\dot{%
a}$, which is further associated with the scalar expansion $\theta= 3\frac{%
\dot{a}}{a}$.

Assume that the universe is described by the homogeneous, isotropic and spatially flat FLRW metric given by, 
\begin{equation}  \label{13}
ds^{2}= -dt^{2}+ a^{2}(t)\delta_{ij}dx^{i}dx^{j}.
\end{equation}
where $a(t)$ is the scale factor of the Universe. Moreover, assume that the
cosmic fluid acquire a bulk viscosity. The energy-momentum tensor can be
written as 
\begin{equation}  \label{14}
T_{\mu \nu}= \rho u_{\mu}u_{\nu} + (p+\Pi) H_{\mu \nu}.
\end{equation}

where in the co-moving coordinates, $u_{\mu} = (1, 0)$, and $H_{\mu \nu} =
g_{\mu \nu}+ u_{\mu} u_{\nu}$. Also, the non-metricity function $Q$ for such
a metric is calculated as $Q=6H^{2}$. By defining the effective pressure as $%
\tilde{p}= p+\Pi$, the Einstein's field equations using the metric and 
\eqref{11} are expressed as

\begin{equation}  \label{15}
3H^2= \frac{f}{4F}-\frac{4\pi}{F} \left[(1+ \widetilde{G})\rho+\widetilde{G} 
\tilde{p}\right],
\end{equation}
and 
\begin{equation}  \label{16}
2\dot{H}+ 3H^2= \frac{f}{4F}-\dfrac{2\dot{F}H}{F}+ \frac{4\pi}{F} \left[(1+ 
\widetilde{G})\rho+(2+\widetilde{G}) \tilde{p}\right].
\end{equation}
Here ($\cdot$)dot represents a derivative with respect to time, besides $F= f_{Q}$, and $8 \pi \widetilde{G}=f_{T}$ denote differentiation with respect to $Q$, and $T$, respectively.

It is assumed that the cold dark matter is highly non-relativistic. As a result, we can consider pressure $p = 0$ and assume that the effect of dark energy on universe evolution is included in the viscous term $\Pi= -\zeta \theta$, which has the dimension of pressure.\newline
We assume the simplest functional form $f(Q, T)= \alpha Q + \beta T$, where $\alpha$ and $\beta$ are constants. Therefore, we get $F= f_{Q}= \alpha$ and $8\pi \widetilde{G}= f_{T}= \beta$. Here, $\beta=0$ reduces to the equivalent case of GR which is well-motivated in literature \cite{Lu/2019}. The considered functional of $f(Q, T)$ was investigated in detail \cite{Yixin/2019}, revealing that the universe experiences an accelerating expansion (naturally $\rho \propto e^{-H_{0}t}$), ending with a de-sitter type evolution.

Solving \eqref{15} and \eqref{16}, we get the following differential
equation. 
\begin{equation}  \label{17}
\dot{H}+ k_{1} H+ k_{2} H^{2}=0.
\end{equation}
where

\begin{equation}  \label{18}
k_{1}= \frac{1}{8\alpha} \left( 15 \beta + 96\pi + 3\beta \left( \dfrac{%
\beta + 16\pi}{3\beta+ 16\pi}\right) \right) \zeta_{0},
\end{equation}

and

\begin{equation}  \label{19}
k_{2}= \frac{1}{8\alpha} \left( 6 \alpha + (15\beta + 96\pi)\zeta_{1} +
\left( \dfrac{\beta+ 16\pi}{3\beta +16\pi}\right) (6\alpha +3\beta
\zeta_{1})\right)
\end{equation}

We consider $\frac{dH}{dt}= H \frac{dH}{d lna}$ and using $a= \frac{1}{1+z}$
(taking $a(t_{0})=1$, we obtained the solution of \eqref{17} as 
\begin{equation}  \label{20}
H(z)= H_{0} \left[ \dfrac{(C (1+z))^{k_{2}}-k_{1}}{C^{k_{2}}-k_{1}}\right] .
\end{equation}

where $H(0)= H_{0}$, the present value of Hubble constant and $C$ is an
integrating constant.
As the Hubble parameter contains model parameters $\zeta_{0}$, $\zeta_{1}$, $\alpha$, $\beta$, and $C$, we try to constrain these with the two different datasets in the following section.

\section{Observational Constraints}\label{sec4}

We studied $f(Q, T)$ gravity with the bulk viscosity formalism in the
preceding sections and obtained an exact solution for the derived field
equations. The solution contains four model parameters $\zeta _{0}$, $\zeta _{1}$, $\alpha $, $\beta $, and a constant $C$. To validate our approach, we must constrain these model parameters with some observational datasets to yield the best fit values for these model parameters. This work has considered two datasets: the observational Hubble datasets consisting of $57$ data points and the newly released Pantheon samples with $1048$ data points. 
To constrain the model parameters, firstly, we employed Python's scipy optimization approach and estimated the global minima for the Hubble function in equation \eqref{20}. The considerable variances in the diagonal elements of the covariance matrix about the parameters are noticed. Then, we used Python's emcee module for the numerical analysis and considered the above estimates as means and a Gaussian prior with a fixed $\sigma = 1.0$ as the dispersion. As a result, we investigated the parameter space surrounding the local minima (or estimates). The method employed with two datasets is explored in greater detail below, with the findings displayed as two-dimensional contour plots with $1-\sigma$ and $2-\sigma$ errors.

\subsection{Hz datasets}
The Hubble parameter, $H=\frac{\dot{a}}{a}$ where $(\cdot)$ represents
derivative with respect to cosmic time $t$. The Hubble parameter is used to analyze the expansion of the universe in observational cosmology. The Hubble parameter can be stated as $H(z)$ $=-\frac{1}{1+z}\frac{dz}{dt}$ as a function of redshift, where $dz$ is obtained through spectroscopic surveys.
In contrast, the measurement of $dt$ gives the model-independent value of the Hubble parameter. Two approaches are extensively used to estimate the value of the $H(z)$ at a specific redshifts. The extraction of $H(z)$ from the line-of-sight BAO data is one, while another is the differential age methodology \cite{h1}-\cite{h19}. A list of revised datasets of 57 points: 31 points from the differential age approach and the other 26 points assessed using BAO and other approaches in
the redshift range $0.07\leqslant z\leqslant 2.42$ is briefly summarized in the reference \cite{sharov}. In addition, we have assumed $H_{0}=69$ Km/s/Mpc for our analysis. The chi-square function is used to find the mean values of the model parameters (equivalent to the maximum likelihood analysis) as

\begin{equation}
\chi _{H}^{2}(\zeta_{0}, \zeta _{1}, \alpha, \beta, C)=\sum\limits_{i=1}^{57}%
\frac{[H_{th}(z_{i}, \zeta _{0},\zeta_{1}, \alpha,\beta, C)-H_{obs}(z_{i})]^{2}}{\sigma _{H(z_{i})}^{2}},  \label{chihz}
\end{equation}

where $H_{th}$ represents the theoretical value of the Hubble parameter and $H_{obs}$ represents the observed value. The standard error in the observed value of the Hubble parameter is represented by $\sigma _{H(z_{i})}$. Here, Table-1 contains the $57$ points of Hubble parameter values $H(z)$ with errors $\sigma _{H}$ from differential age ($31$ points), and BAO and other ($26$ points) approaches, along with references.

\begin{widetext}

\begin{figure}[H]
\centering
\includegraphics[scale=0.45]{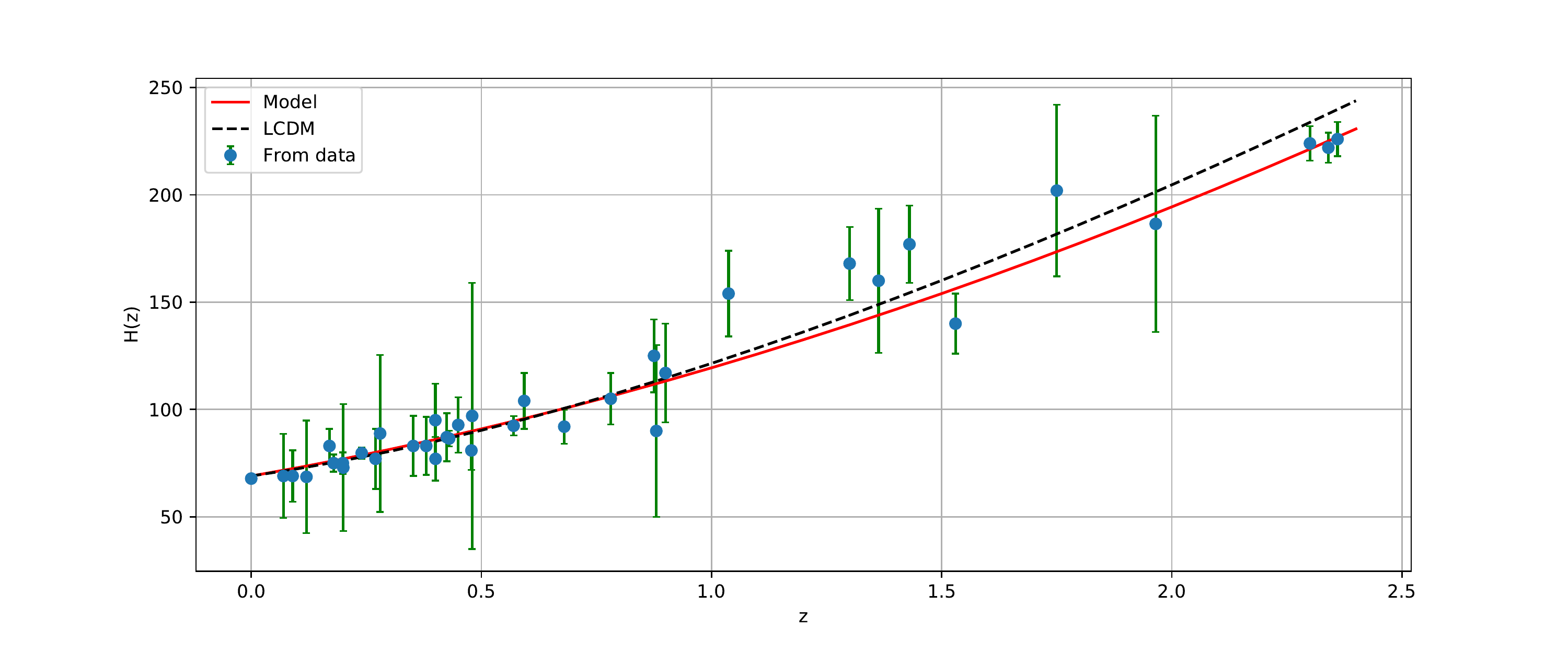}
\caption{The plot shows the evolution of the Hubble function $H(z)$ vs. redshift $z$. The red line shown in the curve is our obtained model. The blue dots shown are the Hubble datasets consisting of $57$ data points with their corresponding error bars, and also the black dashed line depicts the $\Lambda$CDM model with $\Omega_{\Lambda_0}=0.7$ \& $\Omega_{m_0}=0.3$.} \label{Error-Hubble}
\end{figure}
\end{widetext}

\begin{widetext}
\begin{center}
\begin{tabular}{|c|c|c|c|c|c|c|c|}\hline
\multicolumn{8}{|c|}{Table-1: $H(z)$ datasets consisting of 57 data points} \\ \hline
\multicolumn{8}{|c|}{DA method (31 points)}  \\ \hline
$z$ & $H(z)$ & $\sigma _{H}$ & Ref. & $z$ & $H(z)$ & $\sigma _{H}$ & Ref. \\ \hline
$0.070$ & $69$ & $19.6$ & \cite{h1} & $0.4783$ & $80$ & $99$ & \cite{h5} \\ \hline
$0.90$ & $69$ & $12$ & \cite{h2} & $0.480$ & $97$ & $62$ & \cite{h1} \\ \hline
$0.120$ & $68.6$ & $26.2$ & \cite{h1} & $0.593$ & $104$ & $13$ & \cite{h3} \\ \hline
$0.170$ & $83$ & $8$ & \cite{h2} & $0.6797$ & $92$ & $8$ & \cite{h3} \\ \hline
$0.1791$ & $75$ & $4$ & \cite{h3} & $0.7812$ & $105$ & $12$ & \cite{h3} \\ \hline
$0.1993$ & $75$ & $5$ & \cite{h3} & $0.8754$ & $125$ & $17$ & \cite{h3} \\ \hline
$0.200$ & $72.9$ & $29.6$ & \cite{h4} & $0.880$ & $90$ & $40$ & \cite{h1} \\ \hline
$0.270$ & $77$ & $14$ & \cite{h2} & $0.900$ & $117$ & $23$ & \cite{h2} \\ \hline 
$0.280$ & $88.8$ & $36.6$ & \cite{h4} & $1.037$ & $154$ & $20$ & \cite{h3} \\ \hline 
$0.3519$ & $83$ & $14$ & \cite{h3} & $1.300$ & $168$ & $17$ & \cite{h2} \\ \hline 
$0.3802$ & $83$ & $13.5$ & \cite{h5} & $1.363$ & $160$ & $33.6$ & \cite{h7} \\ \hline 
$0.400$ & $95$ & $17$ & \cite{h2} & $1.430$ & $177$ & $18$ & \cite{h2} \\ \hline 
$0.4004$ & $77$ & $10.2$ & \cite{h5} & $1.530$ & $140$ & $14$ & \cite{h2} \\ \hline
$0.4247$ & $87.1$ & $11.2$ & \cite{h5} & $1.750$ & $202$ & $40$ & \cite{h2} \\ \hline
$0.4497$ & $92.8$ & $12.9$ & \cite{h5} & $1.965$ & $186.5$ & $50.4$ & \cite{h7}  \\ \hline
$0.470$ & $89$ & $34$ & \cite{h6} &  &  &  &   \\ \hline
\multicolumn{8}{|c|}{From BAO \& other method (26 points)} \\ \hline
$z$ & $H(z)$ & $\sigma _{H}$ & Ref. & $z$ & $H(z)$ & $\sigma _{H}$ & Ref. \\ \hline
$0.24$ & $79.69$ & $2.99$ & \cite{h8} & $0.52$ & $94.35$ & $2.64$ & \cite{h10} \\ \hline
$0.30$& $81.7$ & $6.22$ & \cite{h9} & $0.56$ & $93.34$ & $2.3$ & \cite{h10} \\ \hline
$0.31$ & $78.18$ & $4.74$ & \cite{h10} & $0.57$ & $87.6$ & $7.8$ & \cite{h14} \\ \hline
$0.34$ & $83.8$ & $3.66$ & \cite{h8} & $0.57$ & $96.8$ & $3.4$ & \cite{h15} \\ \hline
$0.35$ & $82.7$ & $9.1$ & \cite{h11} & $0.59$ & $98.48$ & $3.18$ & \cite{h10} \\ \hline
$0.36$ & $79.94$ & $3.38$ & \cite{h10} & $0.60$ & $87.9$ & $6.1$ & \cite{h13} \\ \hline
$0.38$ & $81.5$ & $1.9$ & \cite{h12} & $0.61$ & $97.3$ & $2.1$ & \cite{h12} \\ \hline
$ 0.40$ & $82.04$ & $2.03$ & \cite{h10} & $0.64$ & $98.82$ & $2.98$ & \cite{h10}  \\ \hline
$0.43$ & $86.45$ & $3.97$ & \cite{h8} & $0.73$ & $97.3$ & $7.0$ & \cite{h13} \\ \hline
$0.44$ & $82.6$ & $7.8$ & \cite{h13} & $2.30$ & $224$ & $8.6$ & \cite{h16} \\ \hline
$0.44$ & $84.81$ & $1.83$ & \cite{h10} & $2.33$ & $224$ & $8$ & \cite{h17} \\ \hline
$0.48$ & $87.79$ & $2.03$ & \cite{h10} & $2.34$ & $222$ & $8.5$ & \cite{h18} \\ \hline
$0.51$ & $90.4$ & $1.9$ & \cite{h12} & $2.36$ & $226$ & $9.3$ & \cite{h19} \\ \hline
\end{tabular}
\end{center}
\end{widetext}

We obtained the best fit values of the model parameters $\zeta _{0}$, $\zeta _{1}$, $\alpha $, $\beta $,  and $C$ as a two dimensional contour plots with $1-\sigma $ and $2-\sigma $ errors in fig. \ref{Cont} using the above stated Hubble datasets consisting of the $57$ points as tabulated in Table-1 \cite{Solanki/2021}. The best fit obtained values are $\zeta_{0}=9.7_{-1.10}^{+1.10}$, $\zeta _{1}=0.046_{-0.039}^{+0.016}$, $\alpha= -1.59_{-0.48}^{+0.79}$, $\beta =-9.38_{-1.30}^{+0.81}$, and $C=7.32_{-1.10}^{+0.92}$. In addition, we have observed the curve fit of the model with error bars for the aforementioned Hubble datasets in fig \ref{Error-Hubble} as well as our resulting model compared to the $\Lambda $CDM model (with $\Omega _{\Lambda 0}=0.7$ and $\Omega _{m0}=0.3$). Our model fits the observational Hubble datasets well, as shown in the plot.

\subsection{Pantheon datasets}

The pantheon sample, which contains $1048$ data points, is the most recently released supernovae type \textit{Ia} dataset. We used this sample \cite{DM/2018} of spectroscopically confirmed SNe \textit{Ia} data points covering the redshift range $0.01<z<2.26$. These data points give the estimation of the distance moduli $\mu _{i}=\mu _{i} ^{obs}$ in the redshift range $0<z_{i}\leq 1.41$. We compare the theoretical $\mu _{i}^{th}$ value and observed $\mu _{i}^{obs}$ value of the distance modulus to find the best fit for our model parameters of the derived model. The distance moduli are the logarithms $\mu _{i}^{th}=\mu(D_{L})=m-M=5\log _{10}(D_{L})+\mu _{0}$ where $m$ and $M$ denote apparent and absolute magnitudes, respectively, and 
$\mu _{0}=5\log \left(H_{0}^{-1}/Mpc\right) +25$ is the marginalised
nuisance parameter. The luminosity distance is considered to be

\begin{eqnarray*}
D_{l}(z) &=&\frac{c(1+z)}{H_{0}}S_{k}\left( H_{0}\int_{0}^{z}\frac{1}{%
H(z^{\ast })}dz^{\ast }\right) , \\
\text{where }S_{k}(x) &=&\left\{ 
\begin{array}{c}
\sinh (x\sqrt{\Omega _{k}})/\Omega _{k}\text{, }\Omega _{k}>0 \\ 
x\text{, \ \ \ \ \ \ \ \ \ \ \ \ \ \ \ \ \ \ \ \ \ \ \ }\Omega _{k}=0 \\ 
\sin x\sqrt{\left\vert \Omega _{k}\right\vert })/\left\vert \Omega
_{k}\right\vert \text{, }\Omega _{k}<0
\end{array}
\right.
\end{eqnarray*}
Here, $\Omega _{k}=0$ (flat space-time). We estimated distance $D_{l}(z)$ and chi square function to measure the difference between the SN \textit{Ia} observational data and the predictions of our model. The $\chi _{SN}^{2}$ function for the Pantheon datasets is taken to be,

\begin{equation}
\chi _{SN}^{2}(\mu _{0}, \zeta _{0},\zeta_{1}, \alpha, \beta, C)=\sum\limits_{i=1}^{1048}\frac{[\mu ^{th}(\mu _{0},z_{i}, \zeta _{0},\zeta _{1}, \alpha, \beta, C)-\mu ^{obs}(z_{i})]^{2}}{\sigma _{\mu
(z_{i})}^{2}},  \label{chisn}
\end{equation}%
$\sigma _{\mu (z_{i})}^{2}$ is the standard error in the observed value.

\begin{widetext}

\begin{figure}[H]
\centering
\includegraphics[scale=0.5]{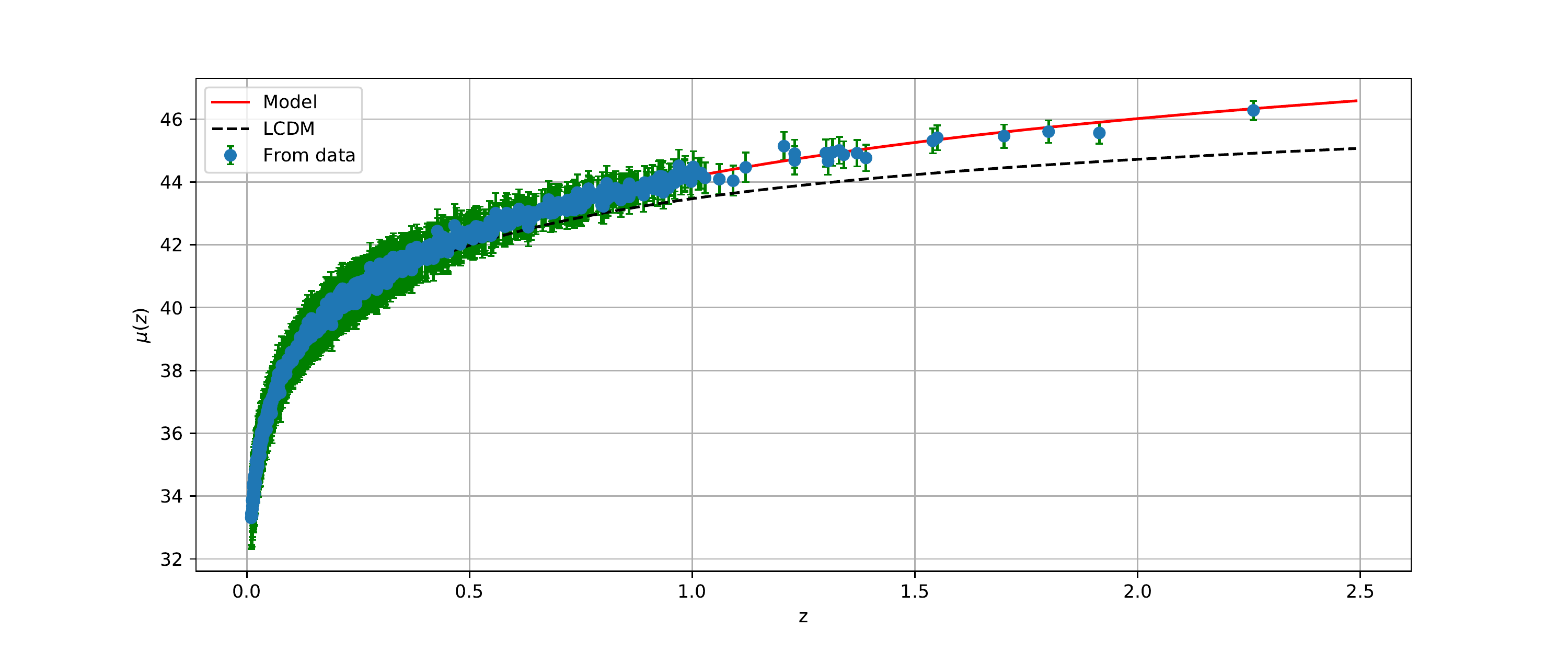}
\caption{The plot depicts the variation of $\mu(z)$ vs. $z$ for our model shown in red line. The black dotted line is the curve for the $\Lambda$CDM model. Both show a nice fit to the Pantheon sample consisting of 1048 data points with the corresponding error bars.} \label{Fig-muz}
\end{figure}

\end{widetext}

\begin{widetext}

\begin{figure}[H]
\centering
\includegraphics[scale=0.75]{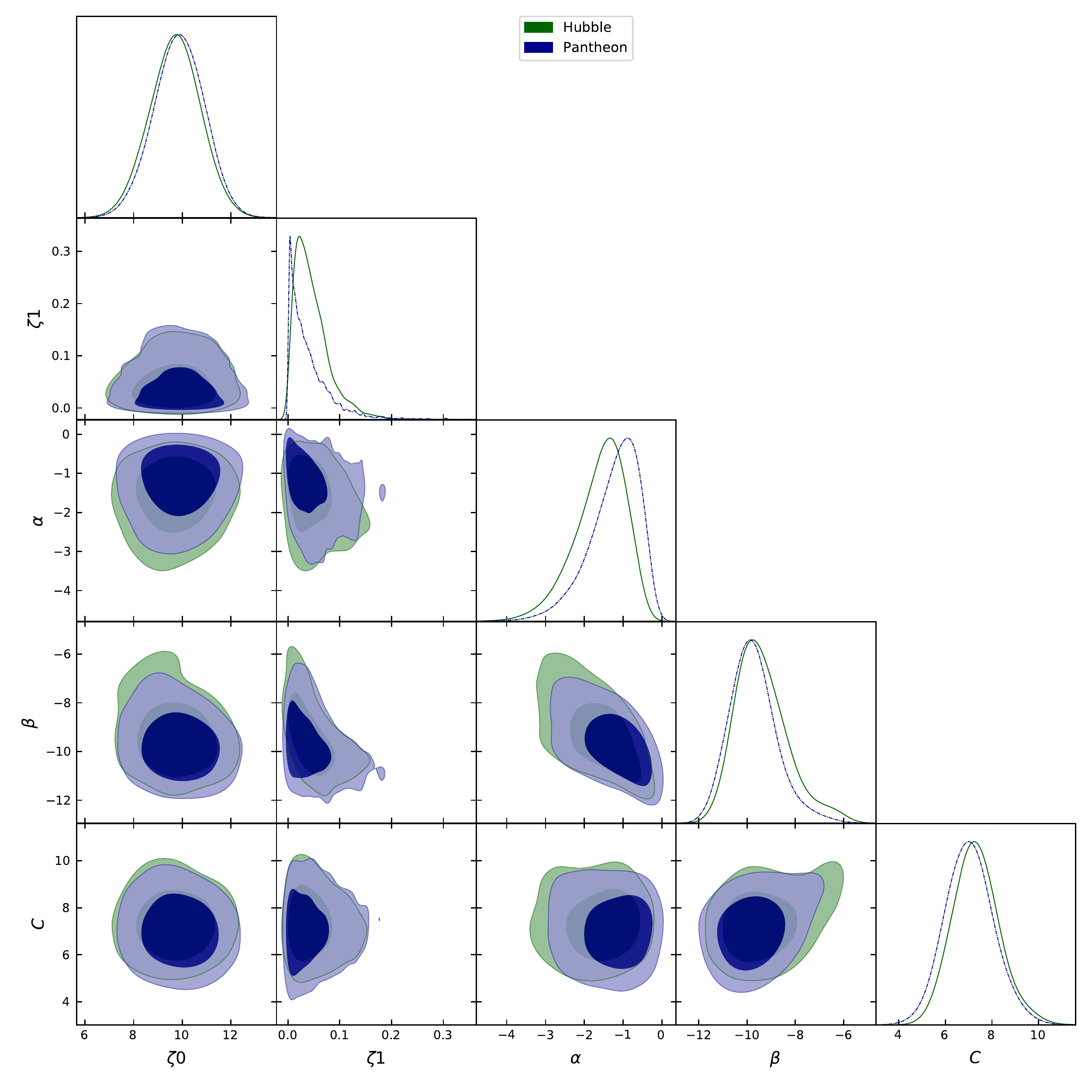}
\caption{The two dimensional contour plots for the model parameters $\zeta _{0}$, $\zeta _{1}$, $\alpha $, $\beta $, and $C$  with $1-\protect \sigma $ and $2-\protect\sigma $ errors. It also includes the best fit values of the model parameters obtained from the Hubble datasets consisting of $57$ points and the Pantheon samples of 1048 points.} \label{Cont}
\end{figure}
\end{widetext}

We have determined the best fit values of the model parameters $\zeta _{0}$, $\zeta _{1}$, $\alpha$, $\beta$, and $C$ using the aforementioned Pantheon datasets as two dimensional contour plots with $1-\sigma $ \& $2-\sigma $ errors in fig. \ref{Cont}. The obtained best fit values are  $\zeta_{0}=9.9_{-1.0}^{+1.0}$, $\zeta _{1}=0.0411_{-0.0420}^{+0.0066}$, $\alpha =-1.22_{-0.40}^{+0.79}$, $\beta =-9.74_{-1.10}^{+0.81}$, and $C=7.04_{-1.10}^{+0.97}$ with $1048$ points of Pantheon datasets. Furthermore, in fig \ref{Fig-muz}, we observed our derived model curve fitting for the mentioned Pantheon datasets along with the error bars and compared to the $\Lambda $CDM model (with $\Omega _{\Lambda 0}=0.7$ and $\Omega _{m0}=0.3$). The plot depicts a good match of our model to the Pantheon observational datasets.

\subsection{BAO datasets}

The analysis of baryonic acoustic oscillations (BAO) deals with the early universe. Baryons and photons are deeply connected in the early universe by Thompson scattering and behave as a single fluid that cannot collapse under gravity and instead oscillate due to the tremendous pressure of photons. These oscillations are named as Baryonic acoustic oscillations (BAO). The characteristic scale of BAO is determined by the sound horizon $r_s$ at the photon decoupling epoch $z_\ast$ given by

\begin{equation*}
r_{s}(z_{\ast })=\frac{c}{\sqrt{3}}\int_{0}^{\frac{1}{1+z_{\ast }}}\frac{da}{
a^{2}H(a)\sqrt{1+(3\Omega _{0b}/4\Omega _{0\gamma })a}},
\end{equation*}
where the quantity $\Omega _{0b}$ is the baryon density and the quantity $\Omega _{0\gamma }$ is the photon density at present time.

The angular diameter distance $D_{A}$ and the Hubble expansion rate $H$ as functions of $z$ are also calculated using the BAO sound horizon scale. If $\triangle \theta $ represents the measured angular separation of the BAO feature in the 2 point correlation function of the galaxy distribution on the sky, and the $\triangle z$ means the measured redshift separation of the BAO feature in the 2 point correlation function along the line of sight then, $\triangle \theta =\frac{r_{s}}{d_{A}(z)}$ where $d_{A}(z)=\int_{0}^{z}\frac{dz^{\prime }}{H(z^{\prime })}$ and $\triangle z=H(z)r_{s}$.

In this work, BAO datasets of $d_{A}(z_{\ast })/D_{V}(z_{BAO})$ from the references \cite{BAO1, BAO2, BAO3, BAO4, BAO5, BAO6} is considered where the photon decoupling redshift is $z_{\ast }\approx 1091$ and $d_{A}(z)$ is the co-moving angular diameter distance. Also, $D_{V}(z)=\left( d_{A}(z)^{2}z/H(z)\right) ^{1/3}$ is the dilation scale. The data used for this analysis is given in the Table-2.

\begin{widetext}
\begin{center}
\begin{tabular}{|c|c|c|c|c|c|c|}
\hline
\multicolumn{7}{|c|}{Table-2: Values of $d_{A}(z_{\ast })/D_{V}(z_{BAO})$
for distinct values of $z_{BAO}$} \\ \hline
$z_{BAO}$ & $0.106$ & $0.2$ & $0.35$ & $0.44$ & $0.6$ & $0.73$ \\ \hline
$\frac{d_{A}(z_{\ast })}{D_{V}(z_{BAO})}$ & $30.95\pm 1.46$ & $17.55\pm 0.60$
& $10.11\pm 0.37$ & $8.44\pm 0.67$ & $6.69\pm 0.33$ & $5.45\pm 0.31$ \\ 
\hline
\end{tabular}
\end{center}

\qquad The chi square function for BAO is given by \cite{BAO6} 
\begin{equation}
\chi _{BAO}^{2}=X^{T}C^{-1}X\,,  \label{chibao}
\end{equation}
where 
\begin{equation*}
X=\left( 
\begin{array}{c}
\frac{d_{A}(z_{\star })}{D_{V}(0.106)}-30.95 \\ 
\frac{d_{A}(z_{\star })}{D_{V}(0.2)}-17.55 \\ 
\frac{d_{A}(z_{\star })}{D_{V}(0.35)}-10.11 \\ 
\frac{d_{A}(z_{\star })}{D_{V}(0.44)}-8.44 \\ 
\frac{d_{A}(z_{\star })}{D_{V}(0.6)}-6.69 \\ 
\frac{d_{A}(z_{\star })}{D_{V}(0.73)}-5.45%
\end{array}%
\right) \,,
\end{equation*}

and $C^{-1}$ is the inverse covariance matrix defined in \cite{BAO6}.

\begin{equation*}
C^{-1}=\left( 
\begin{array}{cccccc}
0.48435 & -0.101383 & -0.164945 & -0.0305703 & -0.097874 & -0.106738 \\ 
-0.101383 & 3.2882 & -2.45497 & -0.0787898 & -0.252254 & -0.2751 \\ 
-0.164945 & -2.454987 & 9.55916 & -0.128187 & -0.410404 & -0.447574 \\ 
-0.0305703 & -0.0787898 & -0.128187 & 2.78728 & -2.75632 & 1.16437 \\ 
-0.097874 & -0.252254 & -0.410404 & -2.75632 & 14.9245 & -7.32441 \\ 
-0.106738 & -0.2751 & -0.447574 & 1.16437 & -7.32441 & 14.5022%
\end{array}%
\right) \,
\end{equation*}
\end{widetext}

\begin{widetext}

\begin{figure}[H]
\centering
\includegraphics[scale=0.75]{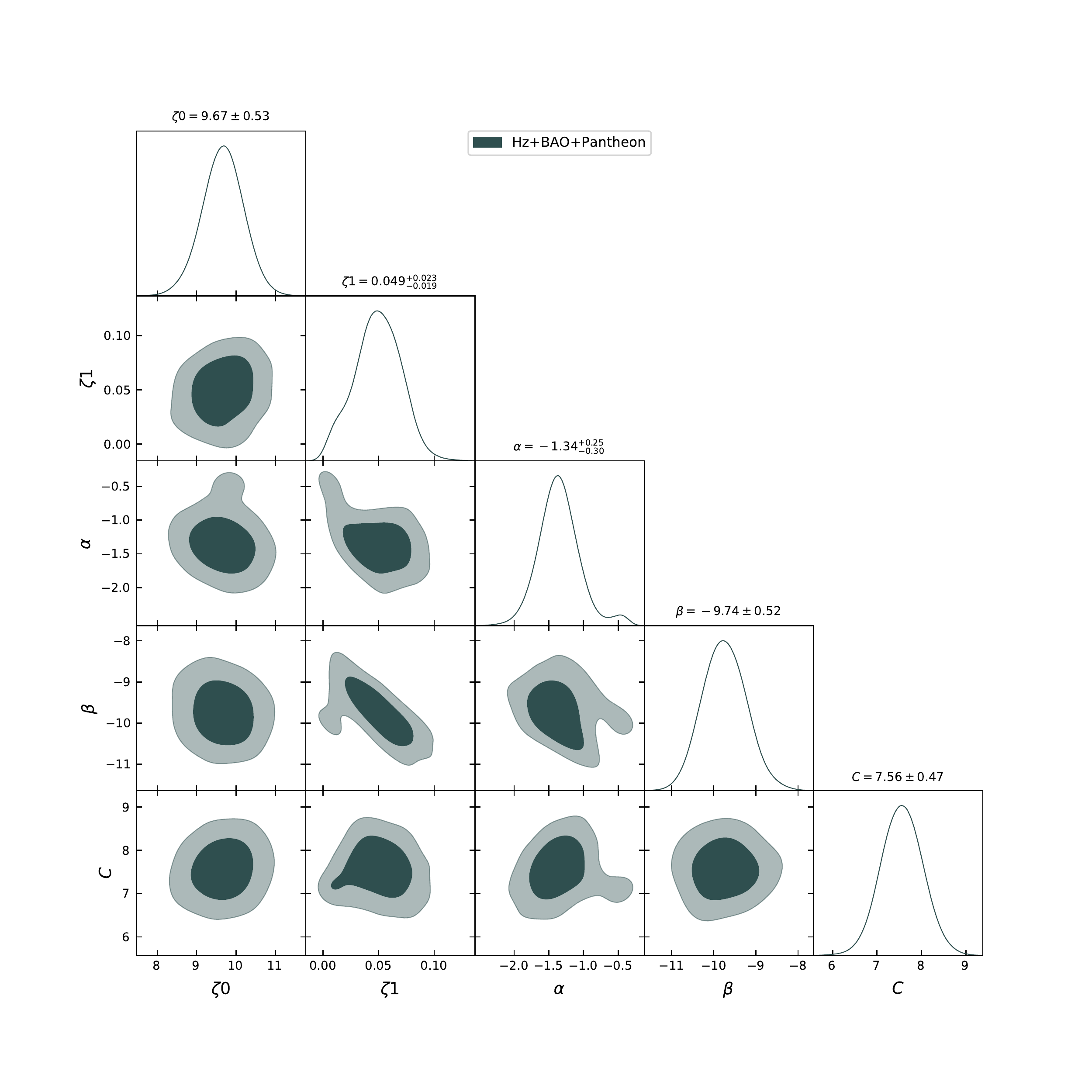}
\caption{The two dimensional contour plots for the model parameters $\zeta _{0}$, $\zeta _{1}$, $\alpha $, $\beta $, and $C$  with $1-\protect \sigma $ and $2-\protect\sigma $ errors. It also includes the best fit values of the model parameters obtained from the combination of Hubble datasets, BAO and Pantheon samples.} \label{Hz+BAO+Pantheon}
\end{figure}

\end{widetext}

Finally, we have taken the combined chi-square for Hubble, Pantheon and BAO datasets and obtained the best fitting values of the model parameters $\zeta _{0}$, $\zeta _{1}$, $\alpha$, $\beta $, and $C$ as shown in fig. \ref{Hz+BAO+Pantheon} as a two dimensional contour plots with $1-\sigma $ and $2-\sigma$ errors. The best fit values for the combined $H(z)+BAO+Pantheon$ datasets are obtained as $\zeta _{0}=9.67_{-0.53}^{+0.53}$, $\zeta _{1}=0.049_{-0.019}^{+0.023}$, $\alpha=-1.34_{-0.30}^{+0.25}$, $\beta =-9.74_{-0.52}^{+0.52}$, and $C=7.56_{-0.47}^{+0.47}$.

We are now fully equipped with all theoretical formulas as well as numerical values of the model parameters and can discuss the physical dynamics of the model. So, the next section is dedicated to the physical dynamics of the other important cosmological parameters.

\section{Cosmological parameters}\label{sec5}

\begin{figure}[H]
\centering
\includegraphics[scale=0.4]{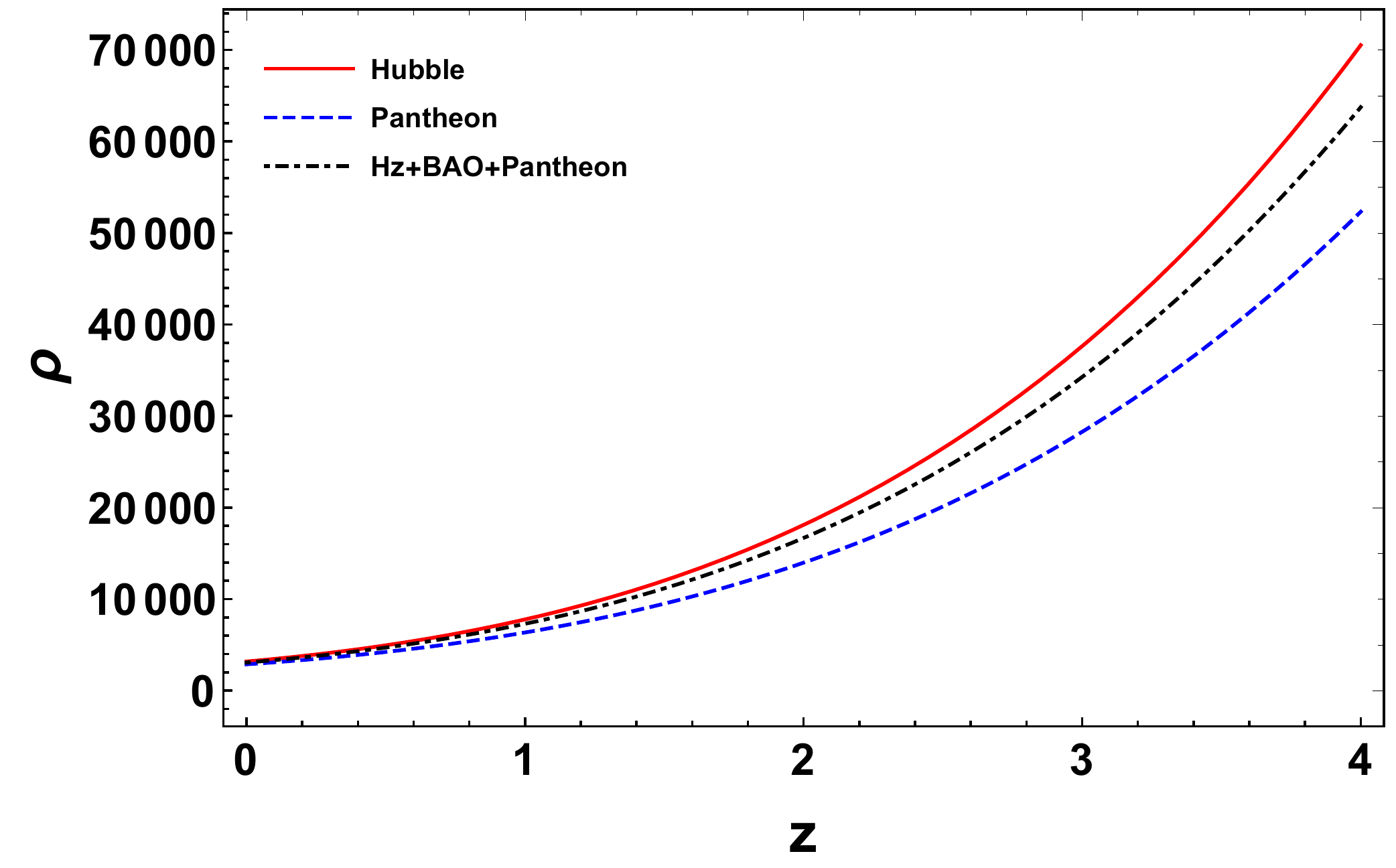}
\caption{The variation of the density parameter versus redshift $z$ for the best
fit values of model parameters $\protect \zeta_{0}$, $\protect\zeta_{1}$, $\protect\alpha$, $\protect\beta$, and $C$ from Hubble data and Pantheon
samples.}
\label{rho}
\end{figure}

\begin{figure}[H]
\centering
\includegraphics[scale=0.4]{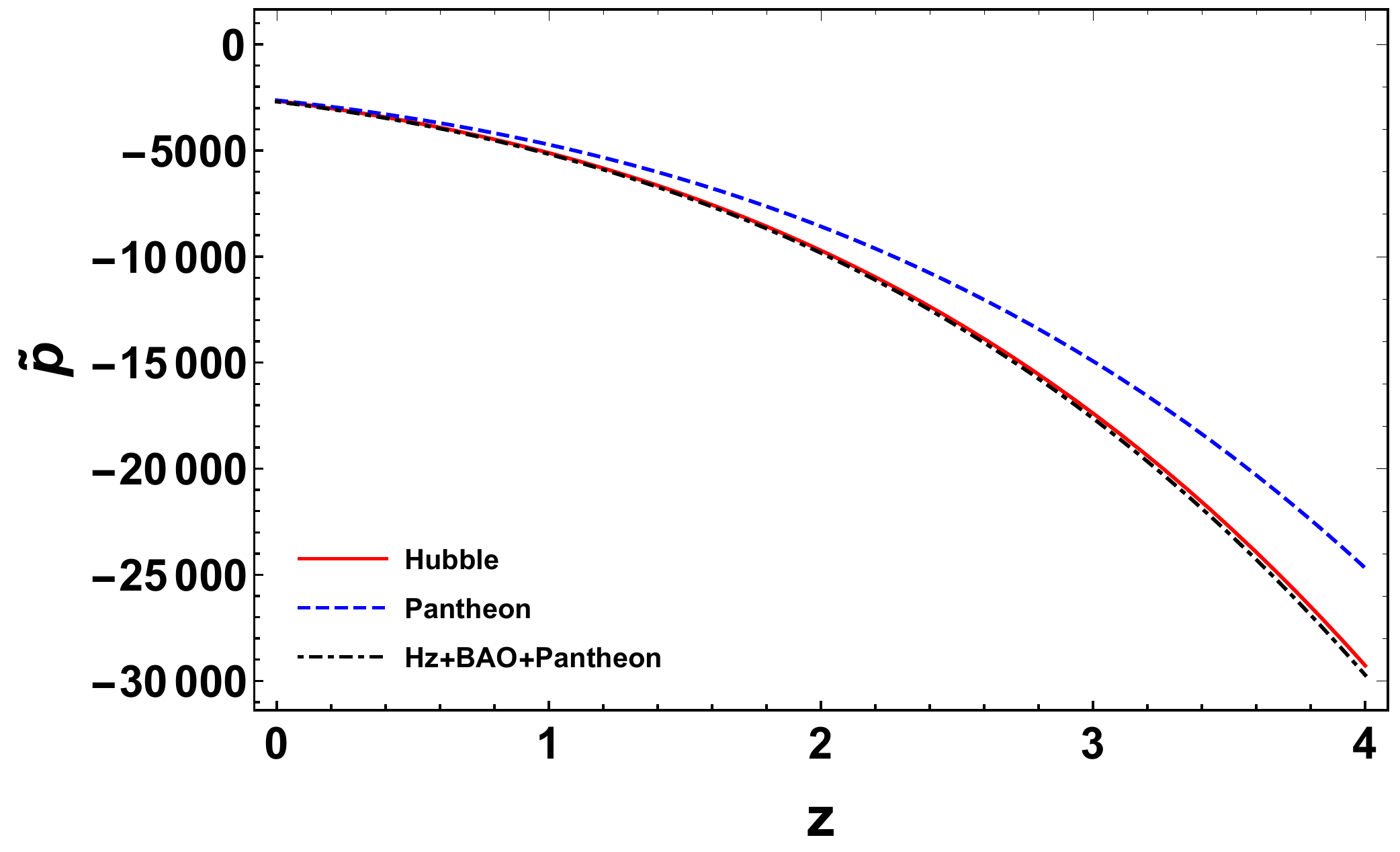}
\caption{The variation of pressure versus redshift $z$ for the best fit
values of model parameters $\protect\zeta_{0}$, $\protect\zeta_{1}$, $\protect\alpha$, $\protect\beta$,  and $C$ from Hubble data and Pantheon samples.}
\label{Pressure}
\end{figure}

The behavior of energy density $\rho$ and effective pressure $\tilde{p}$ is shown in plots \ref{rho} and \ref{Pressure}. It is observed that energy density is an increasing function of $z$, and the effective pressure is strongly negative. The negative pressure is due to the bulk viscosity considered, which indicates the expanding accelerated phase of the universe.
\newline

The deceleration parameter as a function of Hubble parameter $H(z)$ is given
by 
\begin{equation}
q= -1-\frac{\dot{H}}{H^2}.  \label{de}
\end{equation}

\begin{figure}[H]
\centering
\includegraphics[scale=0.4]{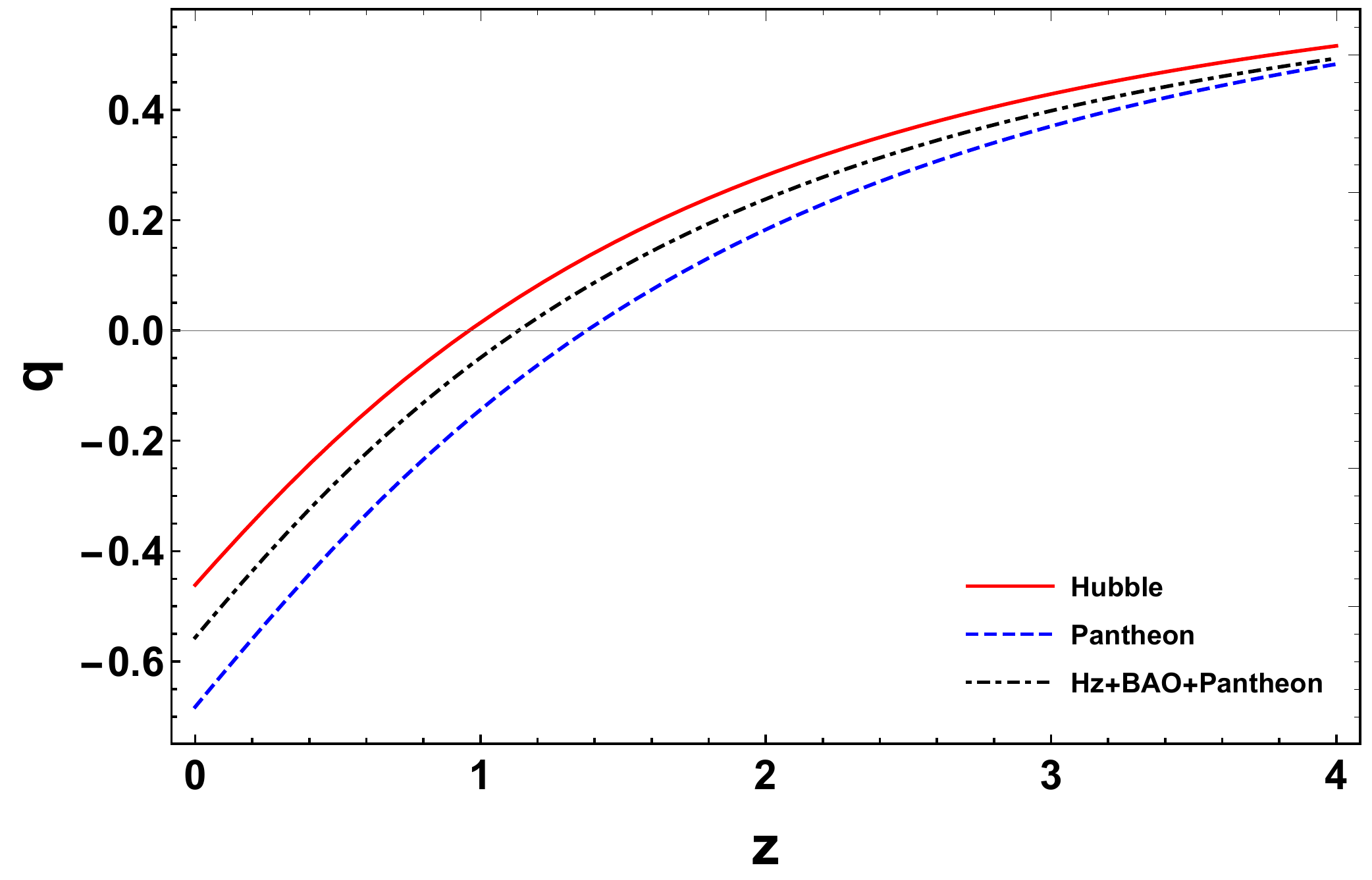}
\caption{The variation of the deceleration parameter versus redshift $z$ for the best fit values of model parameters $\protect\zeta_{0}$, $\protect\zeta_{1}$, $\protect\alpha$, $\protect\beta$,  and $C$ from Hubble data and Pantheon sample.}
\label{Q}
\end{figure}

It describes the rate of expansion and acceleration or deceleration of the universe. If $q>0$, the universe is at a decelerated phase, else $q<0$ corresponds to an accelerated phase. The deceleration parameter $q$ is obtained from the eq. \eqref{de} contains the model parameters $\zeta_{0}$, $\zeta_{1}$, $\alpha$, $\beta$ and $C$. The following plot shows the behavior of $q$ for redshift $z$, explaining the evolution from past to present. Considering the constrained values of model parameters from the two considered datasets, $q$ transit from positive in the past, i.e., early deceleration, to negative at present, indicating the present acceleration in fig. \ref{Q}. The present value of $q$ obtained from Hubble dataset and Pantheon sample is $q_{0}= -0.46^{+1.26}_{-2.93}$ and $q_{0}= -0.68^{+1.43}_{-5.44}$ respectively \cite{Gruber/2014,Mamon/2017,Santos/2016}. In case of combined $Hz+BAO+Pantheon$ data, the value of $q_{0}= -0.55^{+0.80}_{-1.18}$. So, the value obtained from the Pantheon sample is consistent with the $\Lambda$CDM model at $1-\sigma$ level. 

\begin{figure}[H]
\centering
\includegraphics[scale=0.4]{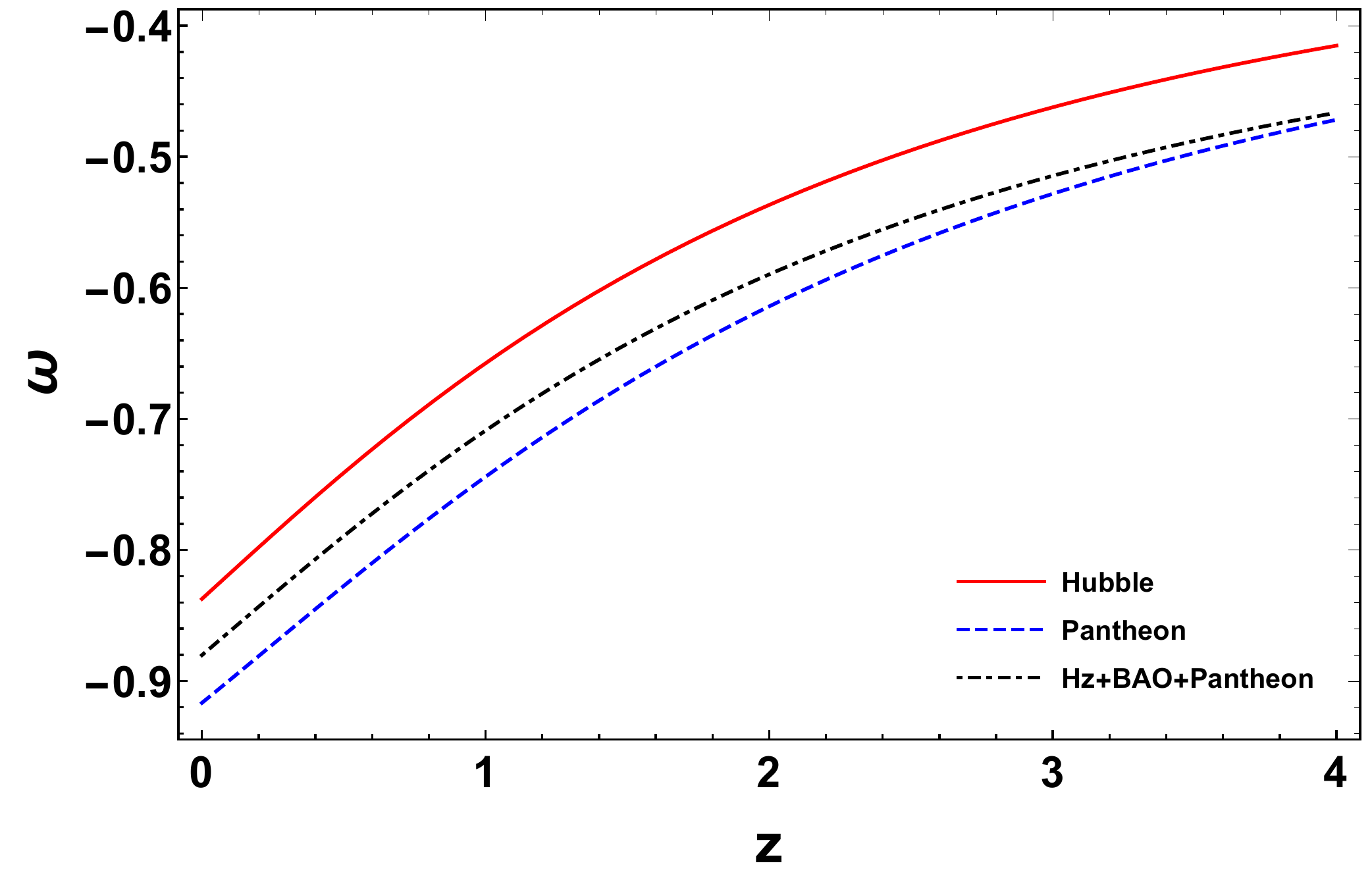}
\caption{The variation of the EoS parameter versus redshift $z$ for the best fit values of model parameters $\protect\zeta_{0}$, $\protect\zeta_{1}$, $\protect\alpha$, $\protect\beta$, and $C$ from Hubble data and Pantheon sample.}
\label{Omega}
\end{figure}

Identifying the value and evolution of the equation of state parameter(EoS) is another attempt to understand the existence of dark energy. The EoS parameter is written as $\omega = \frac{\tilde{p}}{\rho}$, where $\tilde{p}$ is the effective pressure with bulk viscosity and $\rho$ is the energy density. According to observational evidence, the cosmological candidate is a strong candidate for dark energy with $\omega=-1$. However, the accelerating phase of the universe is interpreted when $\omega < -\frac{1}{3}$. This includes $-1 < \omega < 0$ representing the quintessence phase
whereas $\omega$ below $-1$ is known as a phantom regime. \newline
According to constrained values of model parameters $\zeta_{0}$, $\zeta_{1}$, $\alpha$, $\beta$, and $C$ from Hubble data and Pantheon sample, the behavior of the EoS parameter is shown in fig. \ref{Omega}. It is observed that the EoS parameter remains in the quintessence phase supporting the acceleration in the universe. The present value of $\omega$ is obtained as $\omega_{0}= -0.83^{+0.40}_{-0.70}$ and $\omega_{0}= -0.91^{+0.42}_{-0.87}$ \cite{Garza/2019} for Hubble data and Pantheon sample respectively. Further, the present of EoS parameter is $\omega_{0}= -0.88^{+0.24}_{-0.30}$ for the combined $Hz+BAO+Pantheon$ data. The present value of the EoS parameter $\omega$ is in the preferred range i.e, in the neighborhood of -0.8, according to the Planck 2018 results.

\section{Conclusions}\label{sec6}

In this work, we have studied the evolution of the universe in the FLRW framework with the non-relativistic bulk viscous matter in modified $f(Q, T)$ gravity theory. We have considered the bulk viscosity term as $\zeta= \zeta_{0}+\zeta_{1} H$, where $H$ is the Hubble parameter, $\zeta_{0}$ and $\zeta_{1}$ are constants. We have obtained the exact solutions to field equations in the presence of bulk viscosity and assumed the functional form $f(Q, T)= \alpha Q+ \beta T$, where $\alpha$ and $\beta$ are free parameters. Henceforth, the solution obtained is in the form of Hubble parameter containing $\zeta_{0}$, $\zeta_{1}$, $\alpha$, $\beta$ and $C$ as free parameters. Further, we use 57 Hubble data points and 1048 Pantheon samples to constrain the model parameters $\zeta_{0}$, $\zeta_{1}$, $\alpha$, $\beta$ and $C$. As a result, MCMC techniques are used to find the best values for these model parameters. The best values are\newline

Hubble datasets: 
\begin{align*}
\zeta_{0} &=9.7_{-1.1}^{+1.1} & \zeta _{1} &=0.046_{-0.039}^{+0.016} \\
\alpha &= -1.59_{-0.48}^{+0.79} & \beta &=-9.38_{-1.30}^{+0.81} \\
C&=7.32_{-1.10}^{+0.92} & 
\end{align*}

Pantheon datasets: 
\begin{align*}
\zeta_{0}&=9.9_{-1.0}^{+1.0} & \zeta _{1}&=0.0411_{-0.0420}^{+0.0066} \\
\alpha &= -1.22_{-0.40}^{+0.79} & \beta &=-9.74_{-1.10}^{+0.81} \\
C&=7.04_{-1.10}^{+0.97} & 
\end{align*}

Hz+BAO+Pantheon datasets: 
\begin{align*}
\zeta_{0}&=9.67_{-0.53}^{+0.53} & \zeta _{1}&=0.049_{-0.019}^{+0.023} \\
\alpha &= -1.34_{-0.30}^{+0.25} & \beta &=-9.74_{-0.52}^{+0.52} \\
C&=7.56_{-0.47}^{+0.47} & 
\end{align*}

It is found that our model agrees well with the $\Lambda$CDM model as
shown in fig \ref{Error-Hubble} and fig \ref{Fig-muz}.

We investigated the behavior of cosmological parameters for the above best fit-values of model parameters. The density shows the increasing positive behavior, whereas effective pressure is highly negative due to the bulk viscosity. The deceleration parameter depicts a transition from positive in the past to negative in the present showing the current accelerated expansion of the universe. The present value of $q$ is obtained as $q_{0}= -0.46^{+1.26}_{-2.93}$, $q_{0}= -0.68^{+1.43}_{-5.44}$, and $q_{0}= -0.55^{+0.80}_{-1.18}$ for Hubble data, Pantheon samples and the combined $Hz+BAO+Pantheon$ data, respectively. The EoS parameter, on the other hand, is in the quintessence region, indicating that the universe is accelerating. We obtained $\omega_{0}= -0.83^{+0.40}_{-0.70}$, $\omega_{0}= -0.91^{+0.42}_{-0.87}$, and  $\omega_{0}= -0.88^{+0.24}_{-0.30}$ for Hubble data, Pantheon samples and the combined $Hz+BAO+Pantheon$ data, respectively. According to the obtained values of cosmological parameters and behavior, it can be said that the model considered here is more stable with the combined set of data and the bulk viscosity theory is a viable choice for describing the late-time acceleration of the universe in $f(Q, T)$ gravity. Henceforth, it encourages us to investigate the cosmic implications and stability of the newly proposed $f(Q, T)$ gravity in different aspects.

\vspace{0.3cm}

\section*{Acknowledgments}

SA acknowledges CSIR, New Delhi, India for JRF. PKS acknowledges CSIR, New Delhi, India for financial support to carry out the Research
project[No.03(1454)/19/EMR-II Dt.02/08/2019]. 

\end{document}